\documentclass[fleqn,twoside]{article}
\usepackage{espcrc2}

\usepackage{graphicx}

\newcommand{\AmS}{{\protect\the\textfont2
  A\kern-.1667em\lower.5ex\hbox{M}\kern-.125emS}}

\title{Three-nucleon mechanisms in photoreactions}

\author{D.P. Watts\address[GLA]{Department of Physics and Astronomy, University of Glasgow, Glasgow, G12 8QQ,UK.} 
J. Ahrens\address[MAI]{Institut f\"{u}r Kernphysik, Universit\"{a}t Mainz, D-55099 Mainz, Germany}
J.R.M. Annand\addressmark[GLA]
R. Beck\addressmark[MAI]
D. Branford\address[EDI]{Department of Physics and Astronomy, University of Edinburgh, Edinburgh EH9 3JZ, UK}
P. Grabmayr\address[TUB]{Physikalisches Institut, Universit\"{a}t T\"{u}bingen, D-72076 T\"{u}bingen, Germany}
T. Hehl\addressmark[TUB]
J.D. Kellie\addressmark[GLA]
I.J.D. MacGregor\addressmark[GLA]
J.C. McGeorge\addressmark[GLA]
R.O.Owens\addressmark[GLA]}

\begin{document}

\begin{abstract}
The $^{12}$C$(\gamma,ppn)$ reaction has been measured for
  E$_{\gamma}$=150-800 MeV in the first study of this reaction in a
  target heavier than $^3$He. The experimental data are compared to a
  microscopic many body calculation. The model, which predicts that the largest
  contribution to the reaction arises from final state interactions
  following an initial pion production process, overestimates the measured cross sections and there are strong indications that the overestimate arises in this two-step process. The selection of suitable kinematic conditions strongly suppresses this two-step contribution leaving cross sections in which up to half the yield is predicted to arise from the
  absorption of the photon on three interacting nucleons and which agree with the model. The results indicate $(\gamma,3N)$ measurements on nuclei may be a valuable tool for obtaining information on the nuclear three-body interaction.
\vspace{1pc}
\end{abstract}

\maketitle

The study of $(\gamma,3N)$ reactions in nuclei offers a possible new
means of learning about the character of three-nucleon forces (3NF),
which are thought to play a small but essential role in nuclear
structure and reactions\cite{friar}. Since the details of this role
and the relative importance of the different 3NF mechanisms are not
well established it is of interest to examine three-nucleon knockout 
reactions, which can take place via closely related mechanisms, to see 
what light they can cast on the three-nucleon force.

An early indication of the significance of the 3NF came from binding
energy calculations for the few-nucleon systems, which fail to
reproduce the measured binding if only the two-nucleon force is used.
The present position is reviewed in \cite{Gloekle}. The latest
calculations underestimate the binding energy of $^{3}$H and $^{3}$He
by $\sim$0.5-0.9 MeV\cite{Pieper,triton} and of $^4$He by 2.0-4.2 MeV\cite{Pieper,helium} but the inclusion of the
3NF largely removes these discrepancies and also reduces the
disagreements observed in the description of nucleon-deuteron
scattering. Residual discrepancies in the description of polarisation
observables in nucleon-deuteron scattering may be attributable to
additional 3NF mechanisms not presently included in the calculations.

Many 3NF models are based on 2$\pi$ exchange forces and intermediate $\Delta$ excitation but other Feynman
diagrams have been proposed, in particular short range components
involving $N^{*}$ excitation and the exchange of heavier mesons such as the $\rho$ and $\omega$. Recent insights from chiral perturbation theory are helping to guide
the development of 3NF models\cite{friar,epelbaum}. However
experimental input from a range of reactions will be needed to identify the most important diagrams.

To investigate the potential of 3N knockout reactions in the study of
3NFs, it is necessary to determine the contribution made by direct 3N
mechanisms to these reactions and the ease with which they can be
separated from other competing mechanisms. The majority of the
published work on three nucleon knockout from nuclei reports
$(\pi^{+},ppp)$ measurements which have been carried out on
$^{3}$He\cite{lehmann}, $^{4}$He\cite{lehmann,lehmannb} and heavier
targets\cite{kotlinski}. The analysis of these experiments relied on
fitting the data to determine the relative strengths of the contributing reaction mechanisms, in an attempt to separate $(\pi,3N)$ reactions from those where the $(\pi,2N)$ process is preceeded by $(\pi,N)$
scattering or followed by $(N,N^{'})$ scattering. The fits indicate that two step processes involving initial $(\pi,N)$ scattering contribute to the cross section at the $\sim$25\% level. A strong $(\pi,3N)$ contribution is also suggested although significant features in the data were not accounted for. Progress towards identifying the effects of 3NFs requires a theoretical treatment which predicts at least the strength and preferably the details of the $(\pi,3N)$ process.

The use of photons offers advantages over pions as the weak
electromagnetic interaction ensures that the full nuclear interior is
sampled and that initial state interactions, which complicate pion induced reactions, are negligible. The only previous studies of $(\gamma,3N)$ reactions are
$^{3}$He$(\gamma,ppn)$\cite{helium3_gppn} and
$^{12}$C$(\gamma,ppp)$\cite{harty} measurements. For both targets it is concluded that the main contributing mechanism is the 2-step process
involving the initial production of a real pion followed by subsequent
pion reabsorption ($N\pi+ABS$). In \cite{harty} detailed investigation of the
contribution of the $3N$ mechanism to the $^{12}$C$(\gamma,ppp)$ data
was not feasible due to the statistical accuracy obtained and only its
contribution to two-nucleon knockout variables was studied.

With the recent advances in the experimental and theoretical study of
the 3NF, it is timely to explore further what can be learnt about their
nature from $(\gamma,3N)$ measurements on nuclei. As a first step
recent data on the $^{12}$C$(\gamma,ppn)$ reaction, obtained during a
study of $^{12}$C$(\gamma,2N)$ processes\cite{watts}, are examined.  To
gain sensitivity it is important to emphasize the contribution in
which the photon is absorbed by three nucleons ($3N$).

The Valencia model (VM)\cite{valencia,vm_gpi} is a microscopic
many-body model describing photon-nucleus interactions. Its basis is a
construction of the photon self-energy in nuclear matter, which is
related to the cross sections for different photon absorption
mechanisms through Cutkosky rules. A realistic matter distribution for
nuclei is included in this prescription using a local density
approximation. The cross sections for the absorption of the photon by
two nucleons ($2N$), three nucleons ($3N$) and pion
production processes ($N\pi$ and $NN\pi$) are obtained from this theoretical
approach. The $3N$ mechanism includes the effects of both pion and shorter 
range meson exchanges but does not include resonances higher than the $\Delta$.
Both nucleons and pions produced in the absorption process
have quite large reaction cross sections in the nuclear medium and the
model accounts for these final state interactions (FSI) using a
semi-classical Monte Carlo simulation. The Valencia model has been shown to give a quite good general description of $(\gamma,pn)$ reaction on $^{12}$C for $E_{\gamma}$ =110-500 MeV, but overestimates the smaller $(\gamma,pp)$ channel\cite{harty,watts,lamparter,cross}.

The measurements of the $^{12}$C$(\gamma,ppn)$ cross sections reported
here were made using the Glasgow photon-tagging
spectrometer\cite{tagger} at the Mainz 855 MeV electron microtron
MAMI. The tagger produced $\sim$$10^{8}$ tagged photons per second in
the energy range 150-800 MeV, which hit a 4mm thick graphite target.
Products from the resulting photoreactions were detected in two
systems - PiP\cite{pip} and TOF\cite{tof}. PiP is a large ($\sim$ 1sr) 
plastic scintillator hodoscope, which was used to detect
protons in the energy range $\sim$31-290 MeV for polar angles of 78$^{\circ}$-153$^{\circ}$. The proton energy was determined from the energy deposited in the detector. TOF
comprises a large plastic scintillator array which was
used to detect both protons and neutrons in the energy ranges
$\sim$35-190 MeV and 19-640 MeV respectively. The energies of both particle types were determined from the measured time-of-flight. TOF covered polar angles 36$^{\circ}$-142$^{\circ}$ opposite PiP and 16$^{\circ}$-31$^{\circ}$
on the PiP-side of the beam, giving a total solid angle of 0.91 sr.
Reaction timing was obtained from a segmented half-ring of 1mm thick
scintillators ($\Delta$$E_{PiP}$) centred on the target and positioned
on the PiP-side of the beam at a radius of $\sim$11 cm. Separation of
uncharged and charged particles in TOF was carried out using
information from $\Delta$$E_{PiP}$ as well as two further half rings
of 2mm thick scintillator $\Delta$$E_{TOF}$ covering the TOF-side of
the photon beam at radii of $\sim$11 cm and $\sim$30 cm. To remove
ambiguities in this procedure, the two TOF particles were required to
have an opening angle $\geq$60$^{\circ}$. The particle selection and
energy calibration methods employed in the analysis are described
elsewhere\cite{watts}. The setup allows a wide kinematic coverage for
emitted particles with energy resolution of $\sim$6 MeV (FWHM).

The effect of random coincidences in the tagger and neutron detection
in TOF were accounted for using a weight method\cite{weightref}. The
effect of randoms for charged particle detection in PiP and TOF was
negligible due to the requirement of multiple coincidences between the
thick detectors and the $\Delta$$E$ detectors near to the target. The
sources of systematic error in the measured cross sections are
discussed in ref.\cite{harty} and are estimated to be $\pm$12\%. An
overall check of the reliability of the measured cross sections was
provided by a measurement of the D$(\gamma,np)$, $^{12}$C$(\gamma,np)$
and $^{12}$C$(\gamma,pp)$ cross sections, which were all found to
agree\cite{watts} with previous measurements\cite{jenkins_rossi,TYAU}
within the quoted uncertainties.
 
\begin{figure*}
\includegraphics[scale=1.5]{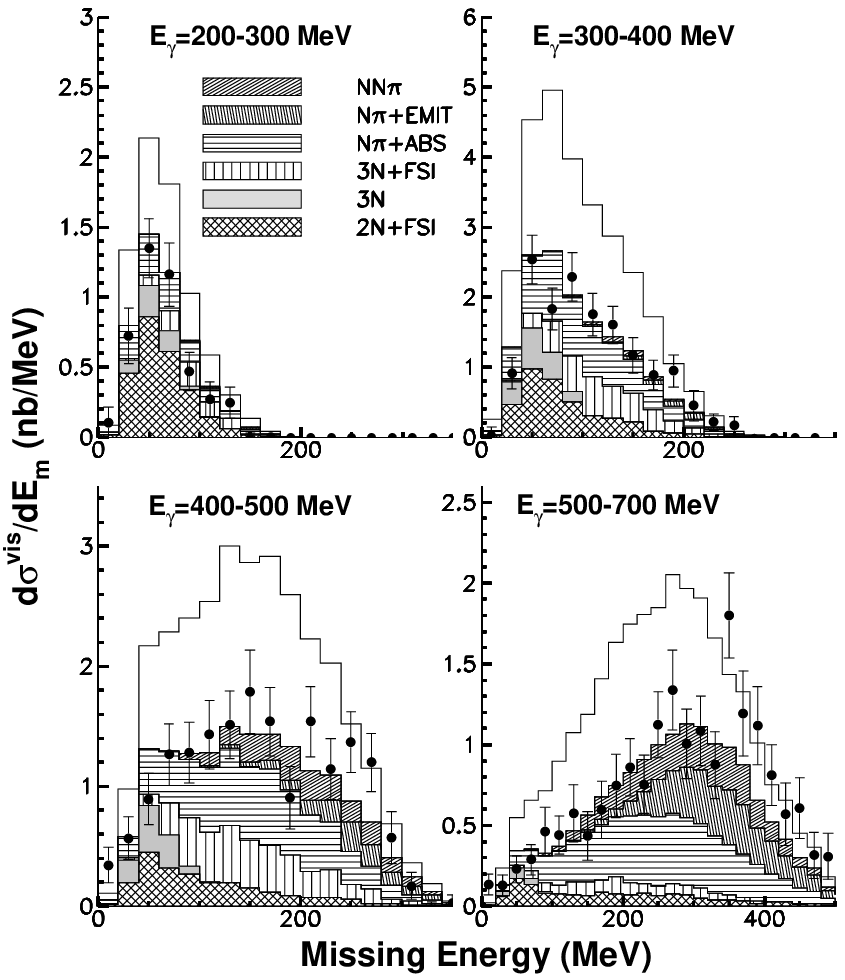}
\caption{\label{fig:em}Three-body missing energy distributions for the $^{12}$C$(\gamma,ppn)$ reaction.  The shaded histograms show the VM predictions for the different mechanisms indicated by the key on the figure. The solid line shows the total cross section without the reduction of the $N\pi+ABS$ prediction (see text).}
\end{figure*}

Fig. \ref{fig:em} shows the measured $^{12}$C$(\gamma,ppn)$ cross
section as a function of the three-body missing energy, defined as
$E^{3B}_m$ $=E_{\gamma}-T_{r} - \sum_{i=1}^{3} T_{i}$, where
$E_{\gamma}$ is the incident photon energy, $T_{i}$ are the kinetic
energies of the three detected nucleons and $T_{r}$ is the (typically
small) energy of the recoiling system which is calculated from its
momentum, ${\bf P_{r}}={\bf P_{\gamma}} - \sum_{i=1}^{3} {\bf P_{i}}$.
The missing energy is the sum of the separation energy $(S)$ for the
reaction and the excitation energy of the (A--3) residual system
($E_{X}$), $E^{3B}_m$=$S$+$E_{X}$.

The experimental data show strength near to $S$ (34.0 MeV)
corresponding to leaving the A--3 system near to its ground state.
There is however significant cross section at higher $E^{3B}_m$
indicating a strong role for processes which leave the A--3 system
highly excited, probably through the emission of more than three
nucleons. Similar features are observed in $(\pi^{+},ppp)$ 
reactions\cite{kotlinski}.

Fig. 1 also shows the result of a VM simulation of the $^{12}$C$(\gamma,ppn)$ reaction including the response of the PiP and TOF detectors. An 8 MeV correction for average nucleon binding is made as described in \cite{harty}. The contribution arising from an intial $N\pi$ process is presented separately for cases where the pion is reabsorbed ($N\pi+ABS$) or when the pion is emitted ($N\pi+EMIT$). Because there is strong evidence from previous experiments\cite{watts,cross} and the 
present results that the strength of the $N\pi+ABS$ process in $^{12}$C is overestimated by the VM model the predictions are presented with the strength of this process reduced by the average factor 0.3 necessary to bring the model into agreement with $^{12}$C$(\gamma,p)$ data\cite{cross} for the backward proton angles sampled in the present measurement. The same factor also produces much closer agreement in comparisons\cite{watts,cross} with $^{12}$C$(\gamma,np)$ data at backward proton angles. Overestimation of the $\pi$ absorption process by the VM by factors of up to $\sim$5 has also been seen for certain kinematic regions in comparisons with $^{12}$C$(\pi^{+},p)$\cite{vm_gpi} and $^{12}$C$(\pi^{+},ppp)$\cite{vm_pi} measurements.

The predicted contributions from different mechanisms (after the
reduction of the $N\pi+ABS$ strength) are separated and shown by the
(stacked) shaded histograms identified in the key in Fig 1. After
the reduction the model gives a good general description of both the
shape and magnitude of the measured data over a wide range of
$E_{\gamma}$ and missing energy. The $3N$ process gives its largest
contribution at low missing energy, but even there generally contributes less than 1/3 of the cross section. Without reduction of the $N\pi+ABS$ process the VM prediction is seen to give a much poorer description of the experimental data and in several cases the predicted $N\pi+ABS$ contribution alone exceeds the measured cross section.

A concerted attempt to minimise the contributions of the absorption
mechanisms other than 3N would employ kinematic cuts based on the
predicted distributions of the nucleons produced by the different
mechanisms. Such predictions are not all available at present. 
In their absence we have made a significant reduction in the $(N\pi+ABS)$ contribution
based on a reconstruction of the invariant mass of the pion, 
which is created in the first step of the process. Neglecting Fermi
motion and nucleon FSI, the invariant mass $(M_{X})$ of the object $X$
in an assumed $\gamma+N \longrightarrow N+X$ reaction can be
reconstructed from the measured 4-vectors of the photon and one of the
detected nucleons
ie. $M_{X_{i}}^{2}=(E_{\gamma}+m_{N_{i}}-E_{N_{i}})^{2}$ - $({\bf
  p}_{\gamma}- {\bf p}_{N_{i}})^{2}$ for i=1,2,3. For on-shell pion production $(M_{X_{i}}/m_{\pi})^{2}$ should be $\sim$1 for the nucleon involved.

\begin{figure*}
\includegraphics[scale=1.3]{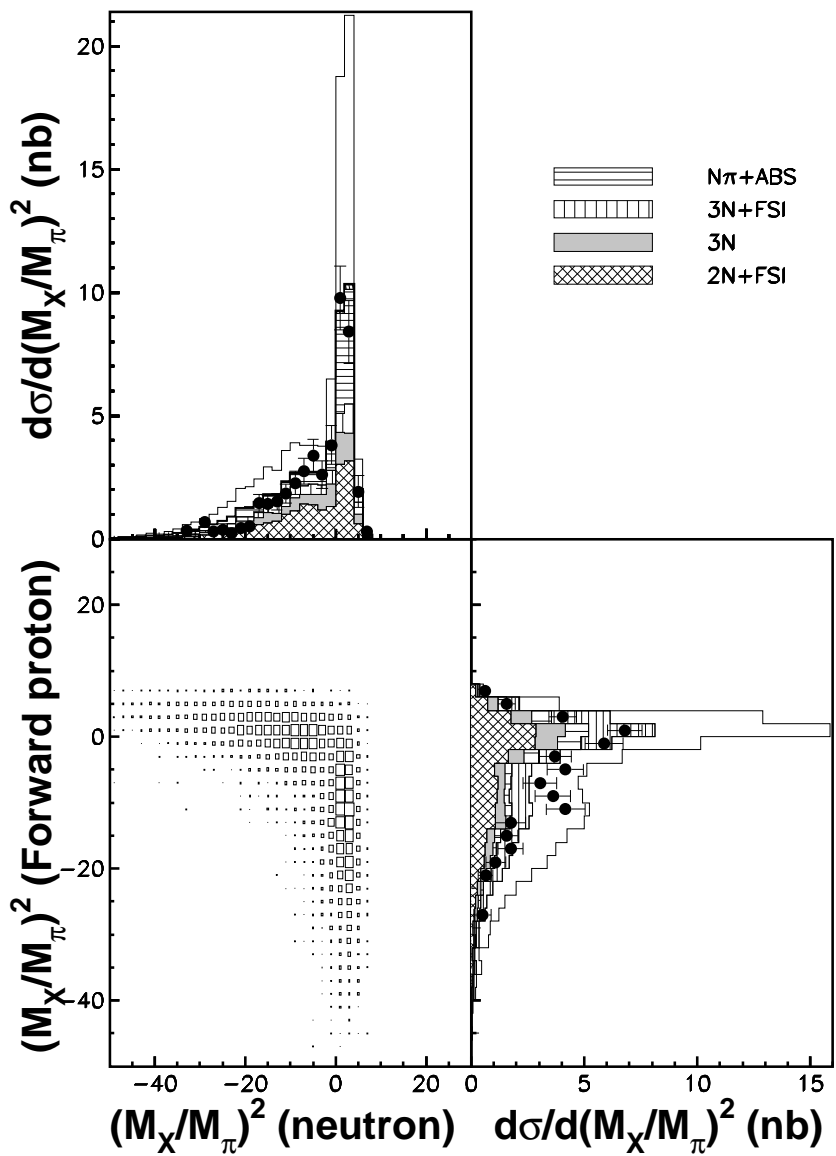}
\caption{\label{fig:mx}Measured and predicted $(M_{X}/M_{\pi})^{2}$ distributions for the $^{12}$C$(\gamma,ppn)$ reaction at $E_{m}^{3B}$$\leq$100 MeV and $E_{\gamma}$=150-500 MeV. Bottom left panel shows the Valencia model predictions of $(M_{X}/M_{\pi})^{2}$ for the neutron versus the forward proton. The projections onto the x and y axes (shaded histograms) are compared with the experimental data in the top left and botton right panels. The solid line shows the total cross section without the reduction of the $N\pi+ABS$ prediction (see text).}
\end{figure*}

Fig. \ref{fig:mx} shows a comparison of the predicted
$(M_{X}/m_{\pi})^{2}$ distributions with the experiment. Both the
model and experiment have been restricted to $E_{m}^{3B}\leq$100 MeV
to include all the strength from the direct $3N$ process but reduce
the contributions from processes involving FSI. The distributions of
$(M_{X}/M_{\pi})^{2}$ are presented for two of the three
detected nucleons, the neutron and the more forward proton (the
backward proton is much less likely to arise from an initial N$\pi$
process due to kinematic restrictions). The experimental data for both 
the proton and neutron exhibit a peak around $m_{\pi} $with a 
tail of strength extending to negative $(M_{X}/M_{\pi})^{2}$ values.

The VM simulations of these distributions are shown in the
2D scatter plot and in the projections. The tendency of the $N\pi+ABS$ contribution to cluster near to $(M_{X}/M_{\pi})^{2}$=1 is clearly visible. The $3N$ and $2N+FSI$ mechanisms are predicted to give
some strength in this region but with relatively more strength in the
negative tail. The relative strength in the peak and tail regions of
the experimental data cannot be well described by the VM without the
reduction in the relative strength of the $N\pi+ABS$ process. 

A further indication of the contributions from different reaction
mechanisms can be obtained from the photon energy dependence of the
cross section which is shown at the top of Fig. 3 for $E_{m}^{3B}\leq$50 MeV,
 $E_{m}^{3B}\leq$100 MeV and $E_{m}^{3B}\geq$100 MeV. The
$E_{m}^{3B}\leq$50 MeV cut restricts the excitation of the residual
nucleus to that expected following the knockout of three $(1p)$ shell
nucleons, whilst the $E_{m}^{3B}\leq$100 MeV cut includes knockout
from all shell combinations and includes most of the $3N$ strength
predicted by the VM (see Fig. \ref{fig:em}). After the reduction of
the $N\pi+ABS$ strength the VM gives a good description of the
experimental data for $E_{m}^{3B}\leq$100 MeV and accounts for the 
observed strong resonance structure in the $\Delta$ region. Good agreement is also observed 
with the $E_{m}^{3B}\geq$100 MeV data for photon energies below $\sim$450 MeV. For higher photon energies the observed excess strength compared to the VM prediction may be attributed to 2$\pi$ production processes which contribute at these energies and are not included in the model.

As an indication of the success to be expected in removing the contributions 
of processes other than 3N from the data, the lower part of Fig. 3 presents the photon energy dependence of the cross section after applying the restriction $(M_{X}/M_{\pi})^{2}\leq$--1.5 to the neutron and more forward proton and thus rejecting the major part of the $N\pi+ABS$ strength for $E_{m}^{3B}\leq$100 MeV (see Fig. 2). The predicted cross sections in this low missing energy region now show little sensitivity to the initial strength of the $N\pi+ABS$ process and are in fair agreement with the data in both shape and magnitude. The agreement for $E_{\gamma}$ below $\sim$250 MeV, where $2N+FSI$ alone is predicted to contribute, gives confidence in the description of this process by the model.

With the above kinematic conditions almost half the cross section for photon energies around 400 MeV is predicted to arise from the absorption of the photon by three interacting
nucleons. Only one other mechanism, $2N$ knockout which is well understood, now competes with the 3N mechanism at low missing energies, which produces a very favourable situation for extracting information on the strength and $E_{\gamma}$ dependence of the 3N component. The different $E_{\gamma}$ dependence predicted for the $3N$ and $2N+FSI$ processes is also beneficial in this regard. Measurements for lighter nuclei to reduce FSI and with wider acceptance detectors and good statistical accuracy would be especially valuable for isolating the contribution of direct 3N absorption and studying the mechanisms responsible.

\begin{figure*}
\includegraphics[scale=1.3]{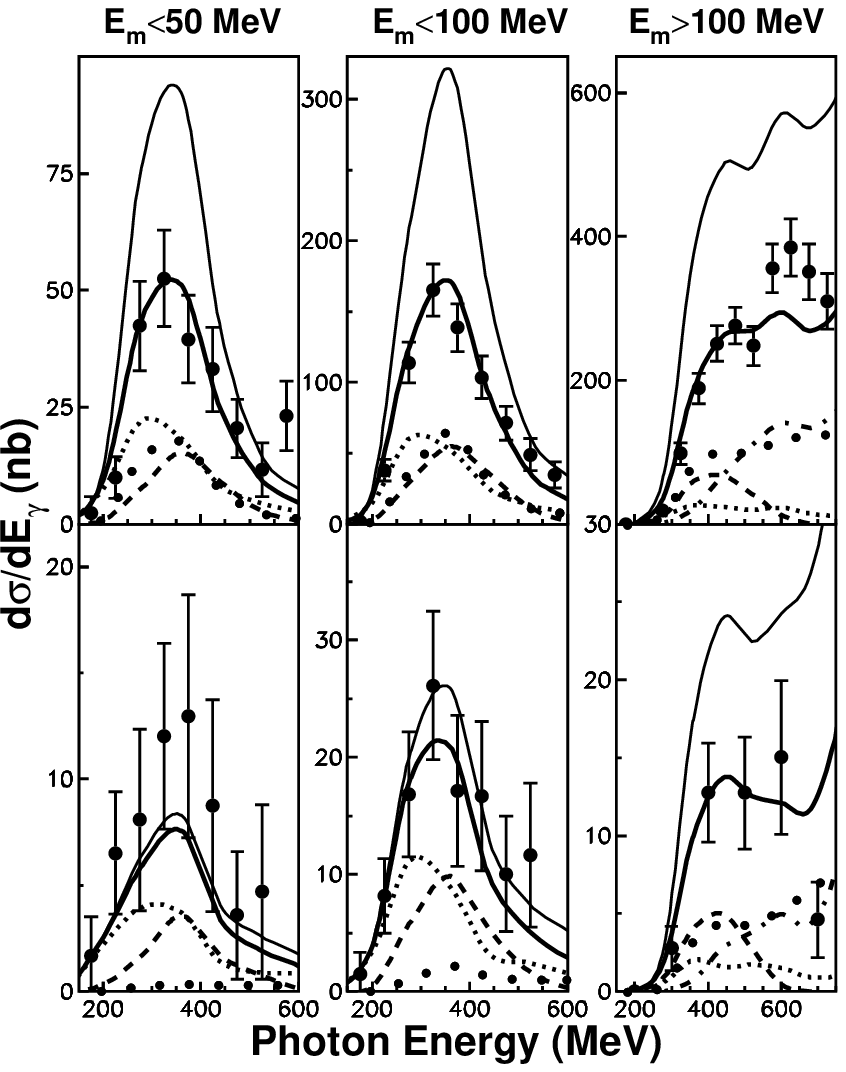}
\caption{\label{fig:egdep}The $^{12}$C$(\gamma,ppn)$ cross section as a function of photon energy presented for three $E_{m}^{3B}$ regions indicated in the figure. The lower figures have the additional cut $(M_{X}/M_{\pi})^{2}\leq$ --1.5 for the neutron and most forward-angle proton. The total prediction of the (modified) Valencia model is shown by the thick solid line and the separate contributions from the $2N+FSI$, $3N$ (with or without FSI), $N\pi$+ABS and $NN\pi/N\pi+EMIT$ process are shown by the short dash, long dash, dotted and dot-dashed lines respectively. The thin solid line shows the total VM without reducing the $N\pi+ABS$ strength (see text).}    
\end{figure*}


\begin{thebibliography}{99}
\bibitem{friar}
J.L. Friar, D. H\"uber, U. van Kolck, Phys. Rev. C 59 (1999) 53.  
\bibitem{Gloekle}
W. Gl\"ockle {\em et. al.}, Nucl. Phys. A 684 (2001) 184c.
\bibitem{Pieper}S.C. Pieper, V.R. Pandharipande, R.B. Wiringa, and J.Carlson, Phys. Rev. C 64 (2001) 014001.
\bibitem{triton} 
J.L. Friar {\em et. al.}, Phys. Lett. B 311 (1993) 4.  
\bibitem{helium} 
J. Carlson, Phys. Rev. C 38 (1998) 1879.; A. Nogga, H. Kamada, W. Gl\"ockle, and B. R. Barrett, Phys. Rev. C 55 (2002) 054003. 
\bibitem{epelbaum}
E. Epelbaum {\em et. al.}, avXiv:nucl-th/0201064
\bibitem{lehmann}
A. Lehmann {\em et. al.}, Phys. Rev. C 55 (1997) 2931.
\bibitem{lehmannb}A. Lehmann {\em et. al.}, Phys. Rev. C 56 (1997) 1872.
\bibitem{kotlinski}B. Kotlinski {\em et. al.}, Eur. Phys. J. A 1 (1998) 435.
\bibitem{helium3_gppn}G. Audit {\em et al}, Nucl. Phys. A 614 (1997) 461; Phys. Lett. B 312 (1993) 57; Phys. Rev. C 44 (1991) R575; A. J. Sarty {\em et al}, Phys. Rev. C 47 (1993) 459.
\bibitem{harty}P.D. Harty {\em et. al.}, Phys. Rev. C 57 (1998) 123.
\bibitem{watts}
D.P. Watts {\em et. al.}, Phys. Rev. C 62 (2000) 014616.  
\bibitem{valencia}R.C. Carrasco and E. Oset, Nucl. Phys. A 536 (1992) 445; {\em ibid} A 570 (1994) 701.  
\bibitem{vm_gpi}R.C. Carrasco,
  E. Oset, and L.L. Salcedo, Nucl. Phys. A 541 (1992) 585.
\bibitem{vm_pi} M.J. Vicente-Vacas and E. Oset, Nucl. Phys. A 568 (1994) 855.  
\bibitem{lamparter}T. Lamparter {\em et. al.}, Z. Phys. A 355 (1996) 1.  
\bibitem{cross}G.E. Cross {\em et. al.}, Nucl. Phys. A 593 (1995) 463.   
\bibitem{tagger}S.J Hall {\em et. al.}, Nucl. Inst. and Meth. A 368 (1996) 698; I. Anthony {\em et. al.}, {\em ibid} A 301 (1991) 230. 
\bibitem{pip}I.J.D. MacGregor {\em et. al.}, Nucl. Inst. and Meth. A 382 (1996) 479.  
\bibitem{tof}P. Grabmayr {\em et. al.},
  Nucl. Inst. and Meth. A 402 (1998) 85. 
\bibitem{weightref}J. C. McGeorge (private communication); D. Branford {\em et. al.}, Phys. Rev. C 61 (1999) 014603.  
\bibitem{jenkins_rossi}D.A. Jenkins, P.T.
  Debevec, and P.D. Harty, Phys. Rev. C 50 (1994) 74.
\bibitem {TYAU}I.J.D. MacGregor {\em et al.}, Phys. Rev. Lett. 80 (1998) 245.  \end{thebibliography}
\end{document}